\documentstyle[12pt]{ioplppt}
\newcommand{\myvec}[1]{\mbox{\boldmath{$#1$}}}

\begin{document}

\title{A solitary-wave representation of turbulence 
 	in the physical-plus-eddy space}
\author{S.Tsug\'e\\
  Tsukuba Advanced Research Alliance, University of Tsukuba,\\
  Tsukuba,305 Japan\ftnote{1}{present address : 790-3 Tohigashi, Tsukuba
    300-26 Japan}\ftnote{0}{e-mail address : shunt@tara.tsukuba.ac.jp}
  }
\begin{abstract}
  A unique form of 
turbulent-transport equations is derived based on first principles.The
role of nonequilibrium statistical mechanics employed to describe the
phenomenology is that it enables to single out the unique form
consistent with master equation of Liouville, a prerequisite not met
with existing equations for turbulence modeling.The equation is
variable-separated to yield a Navier-Stokes equation in
6D(physical-plus-eddy) space with homogeneous boundary
conditions.Turbulent transports such as Reynolds' stress are calculated
using a solution of this equation; a solitary-wave function.Satisfactory
agreement is observed with existing experiment for mixing shear layer of
incompressible flows although no empirical constants to fit with data
are involved.
\end{abstract}
\section{Introduction}
First principle bases of phenomenologies of fluid dynamics and
thermodynamics are due, respectively, to
Chapman-Enskog\cite{b1},\cite{b2} and
Prigogine\cite{b3}, who showed the validity of Navier-Stokes equation
and the equality expression of the second law of thermodynamics on
common basis of the Boltzmann equation expanded to the first order
deviation from equilibrium.

It was, however, an open question whether these equations still hold
for turbulent flows where the dependent variables are stochastic and
fractal, therefore not differentiable\cite{b4}. This problem 
was solved\cite{b5}
using the microscopic density\cite{b6}, namely, unaveraged Boltzmann
function, leading to rederivation of the Navier-Stokes equation
written in terms of instantaneous quantities without assuming any
statistical concepts like local equilibrium, or first order deviation
form it. Thus the equations currently employed for the direct numerical
simulation(DNS) has acquired the first principle basis. 

Then a question will arise as to whether the alternative
methodology of computational fluid dynamics, namely, Reynolds-average
formalism can be founded on first principles as well.

In contrast with DNS founded on the microscopic density(the Klimontovich 
formalism) its Reynolds-average counterpart should be based on its
averaged version, namely, the Boltzmann
function. Structure of this {\it Boltzmann formalism}, so to say, is
 shown to be identical with what is called the BBGKY hierarchy theory of
nonequilibrium statistical mechanics\cite{b7}.

Each of the two formalisms has advantages as well as disadvantages
from computational viewpoints. The most serious disadvantage of the
K-formalism lies in the fact that it deals with fractal quantities which
are selfsimilarly rugged to a length scale as small as the Kolmogorov
scale\cite{b8}. It means that if the 3-D Navier-Stokes equation is
to be solved by a finite difference method, the grid size be smaller than
this length, namely of $O(R^{-3/4})$ ($R$ ; Reynolds number). It
requires the computer memory size growing with $R^{9/4}$(the small eddy 
difficulty). For this reason, the consensus upper limit of 
applicability of the current DNS falls short of $R\sim 10^4$. To be able to apply 
the DNS for practical design
of transport vehicles($R\sim10^7$) we would have to have a computer
with memory size greater by the factor of $(10^7/10^4)^{9/4}\sim10^7$
compared to
the currently available ones. This situation is not changed even when
one employs spectral methods to avoid the conceptual difficulty of
having to work  with differential equations. For, then, the number of the
Fourier modes to be taken into account increases at the same rate.

The B-formalism, in contrast, is free from such difficulty owing to
the statistical average taken in the process of generating
distribution functions at the expense of dealing with multitude of
such functions, infinite in number. The statistical average as meant
here is either an ensemble average over repeated experiments, or an
average over time that is long enough for
the fractal ruggedness to be smoothed out, yet is short enough for
fluid-dynamic unsteadiness, such as shedding period of K\'arm\'an
vortices or aerodynamic flutter to be discernible. The ergodic theorem
warrants their identity.

The objective of this paper is to demonstrate that the Reynolds
average regime of the computational fluid dynamics can be founded on a 
firm first principle ground through the following processes :
\begin{itemize}
\item[i)] 
  Demonstration of how a closed set of kinetic equations is obtained out of an
  infinite chain of them of the B-formalism.
\item[ii)]
  Taking fluid moments of the kinetic equation for two-point 
  fluctuation-correlations that is unique from the viewpoint of first
  principles.
\item[iii)]
  Rederivation of the whole set of equations thus obtained from
  phenomenological equations being used for DNS, namely, eliminating 
  the process via statistical mechanics. 
\end{itemize}
\section{The Boltzmann and Klimontovich formalisms : A review}
Any statistical theory rests on the axiom that a field quantity
$\underline{f}(z)$ describing stochastic and possibly fractal
  physical phenomenon is equivalent 
to a set of quantities that are smooth, deterministic, and infinite in 
number, 
\begin{eqnarray}\label{eq1}
 \underline{f} \equiv
  \left\{
    \begin{array}{cc}
      f(=\overline{\underline{f}}) &  \\
      {}&\\
      \overline{f^{'} \hat{f}^{'}} & 
      \hspace{5mm}(f^{'} \equiv \underline{f} -f) \\
      {}&\\
      \overline{f^{'} \hat{f}^{'} \tilde{f}^{'}} & \\
      \dots & \\
      \dots &
    \end{array}
  \right.
\end{eqnarray}
where overbar stands for the statistical average as defined in the  
preceding section, and 
$\underline{\hat{f}} = \underline{f} (\widehat{z})$ is the same quantity at a different
point $\widehat{z}$.

The Klimontovich formalism as defined here is a formalism where
$\underline{f}$ is identified with the microscopic density \cite{b6}
\begin{equation}\label{eq2}
  \underline{f}(z)= \sum_{1 \le n \le N} \delta(z-z^{(n)}(t))
\end{equation}
In this expression $z \equiv (\myvec{x,v})$ is a point in the
phase($\mu$-) space, $z^{(n)}(t)$ is the locus of $n$-th molecule in
this space, $N$ is the total number of molecules under consideration,
and $\delta$ denotes the six-dimensional delta function.

It has been shown \cite{b9} that the equation governing $\underline{f}$ is
the {\em unaveraged} Boltzmann equation;
\begin{equation}\label{eq3}
  B[\underline{f}] \equiv (\frac{\partial}{\partial t}+
  \myvec{v} \cdot \frac{\partial}{\partial \myvec{x}})\underline{f} -
  J(z|\widehat{z})[\underline{f} \ 	\underline{\hat{f}}] = 0  
\end{equation}
with $J$ denoting the classical collision integral acting on molecule
$\widehat{z}$ .
%
The key issue here is that Eq.(\ref{eq3}) is a {\it deterministic} equation
of continuity in the $\mu$-space, free from any statistical
concepts\cite{b9}. 

The Boltzmann formalism, on the other hand, deals with quantities to
have appeared on the r.h.s. of (\ref{eq1}), subject to statistical average,
namely, the Boltzmann function
\begin{equation}\label{eq5}
  f(z) \equiv \overline{\underline{f}}
\end{equation}
and the correlation functions of consecutive hierarchies
\begin{equation}\label{eq6}
  \left.
    \begin{array}{rcccl}
      \psi(z,\widehat{z}) &\equiv & \overline{f^{'} \hat{f}^{'}}
     &=& \overline{\underline{f} \ \underline{\hat{f}}} - f \hat{f} \\
      {}\\
      \psi(z,\widehat{z},\widetilde{z}) &\equiv& \overline{f^{'} \hat{f}^{'}
 \tilde{f}^{'}} &=& \overline{\underline{f} \  
\underline{\hat{f}} \ \underline{\tilde{f}}} - f \psi(\widehat{z}, \widetilde{z}) - \hat{f}
      \psi(\widetilde{z},z) - \tilde{f} \psi(z,\widehat{z}) \\
      \cdots\cdots
    \end{array}
  \right\}
\end{equation}

Klimontovich variables \cite{b5} such as instantaneous gas density
$\underline{\rho}$ and center-of-mass velocity $\underline{\myvec{u}}$ of the
$N$-particle system are given by 
\begin{equation}\label{eq9}
  \underline{\rho} = m \hspace{-1mm}\int^{\infty}_{-\infty}  \underline{f} \;d \myvec{v} = m \sum_{1 \le n \le N}
  \delta(\myvec{x}- \myvec{x}^{(n)}(t))\hspace{5mm}
(m : \mbox{mass of a molecule})
\end{equation}
\begin{equation}\label{eq10}
  \underline{\rho} \ \underline{\myvec{u}} = m \hspace{-1mm}\int^{\infty}_{-\infty}  \myvec{v} \underline{f} 
   \;d \myvec{v} =m \sum_{1 \le n \le N} \myvec{v}^{(n)}(t)\; \delta(\myvec{x}-
  \myvec{x}^{(n)}(t)) 
\end{equation}
The Reynolds-averaged version of those fluid variables are generated by
taking average of expressions (\ref{eq9}) and (\ref{eq10})
\begin{equation}\label{eq7}
\rho= m \hspace{-1mm}\int^{\infty}_{-\infty}  f \:d \myvec{v}
\end{equation}
\begin{equation}\label{eq8}
  \overline{\underline{\rho}\: \underline{\myvec{u}}} =
\rho\:\myvec{u} + \overline{\rho'\myvec{u}'}
= m \hspace{-1mm}\int^{\infty}_{-\infty}  \myvec{v} f \:d \myvec{v}
\end{equation}

Two-point correlation $\psi(z,\widehat{z})$ consists of two parts;
short-range part due to direct molecular collisions that 
is irrelevant to turbulence and long-range part attributable to
turbulence correlations. The latter part is expanded in a
double series of Hermite polynomials\cite{b10} where all the turbulent
transport terms including Reynolds' stress appear as expansion
coefficients. 

A set of equations governing those quantities standing on the
l.h.s. of (\ref{eq6}) is generated from Eq.(\ref{eq3}) by taking
moments and averaging. This procedure is not unique; in general there
are infinite ways of constructing such moment equations. This
arbitrariness is eliminated by invoking a postulate that
{\em the whole system be consistent with Liouville's 
  equation}(the equation of continuity of $N$-particle probability
  density in 6$N$-dimensional space; the {\it master} equation  that is 
  universally valid), or its corollary that {\em the whole set of equations 
  be identical with those of the BBGKY theory at each level
  of hierarchies}.
The only difference is that the {\it BBGKY\/} generates destribution
functions in the direction of descending number of molecules through a
series of integrations starting from $N \sim O(10^{20})$, whereas here are defined the same functions in ascending number of molecules. The kinematical information missing in the latter approach is identical in analogy with the fact that we cannot predict the functional form of $g(x,y)$ out of $\displaystyle f(x)=\int^{\infty}_{-\infty} g(x,y) dy$ . This is why we need the postulate.

The averaging process consistent with the postulate has led to following set of equations \cite{b9}\; :\\

$\bullet$ 1 - {\it particle level\/}
\begin{equation}\label{eq11}
  \overline{B[\underline{f}]} = 0: \hspace{3mm}
    (\frac{\partial}{\partial t} + \myvec{v} \cdot \frac{\partial}{\partial
      \myvec{x}})f - J(z|\widehat{z}) [f \hat{f} + \psi(z, \widehat{z})] =0
\end{equation}

$\bullet$ 2 - {\it particle level\/}
\begin{equation}\label{eq12}
\left.
	\begin{array}{l}
\overline{\hat{f}^{'} B[\underline{f}] + f^{'} \widehat{B} [\underline{\hat{f}}]}
 = 0 \;\;\;\;,\;\;\;( \widehat{B} [\;\;]\equiv \{B [\;\;]\}_{z\rightarrow\widehat{z}} )  :\\
\vspace{-2mm}\\
\displaystyle  (\frac{\partial}{\partial t} + \myvec{v} \cdot 
  \frac{\partial}{\partial \myvec{x}} + \widehat{\myvec{v}} \cdot 
  \frac{\partial}{\partial \widehat{\myvec{x}}})\: \psi(z,\widehat{z}) 
- J (z| \widetilde{z}) \:[\:f \psi(\widehat{z}, \widetilde{z})
+ \tilde{f} \psi (z,\widehat{z}) + \\
\vspace{-2mm}\\
\hspace{10mm}  \psi(z,\widehat{z},\widetilde{z})\:] -
  J(\widehat{z} |\widetilde{z})\:[\: \hat{f} \psi(z,\widetilde{z}) + \tilde{f} \psi(z,
  \widehat{z}) + \psi(z,\widehat{z},\widetilde{z})\:] = 0 
\end{array}\right\}
\end{equation}

$\bullet$ 3 - {\it particle level\/}
\begin{equation}\label{eq13}
  \overline{\hat{f}^{'} \tilde{f}^{'} B [\underline{f}] + f^{'}
    \hat{f}^{'} \widetilde{B} [\underline{\tilde{f}}] + \tilde{f}^{'} f^{'}
    \widehat{B}[\underline{\hat{f}}]} = 0 
\end{equation}
This system constitutes a chain of equations for the set of infinite
number of variables $[f(z), \psi(z,\widehat{z}),
\psi(z,\widehat{z},\widetilde{z}),\cdots]$.

The issue that is most crucial to the quality of the proposed approach  
is the closure condition. It
is how to truncate the infinite chain of equations to make the system
tractable without violating physical soundness.

Early stage of development along this line has employed the following
condition of {\it tertiary chaos} \cite{b10}, \cite{b11}
\begin{equation}\label{eq14}
  \psi(z,\widehat{z},\widetilde{z}) = 0
\end{equation}
a condition next to the simplest one known as Boltzmann's (binary)
molecular chaos hypothesis; $\psi(z,\widehat{z})=0$.

Eq.(\ref{eq12}) under this closure condition is
investigeted in some depth: It is shown that assumption (\ref{eq14}) 
allows Eq.(\ref{eq12}) for separating variables into those for respective 
particles, thereby its fluid moment
equation leads to linearized Navier-Stokes equation (the Orr-Sommerfeld
equation, in particular.). It is also seen that this closure 
gives satisfactory description only for weak turbulence where the
nonlinearity in turbulent intensity does not play major roles. Later, 
it has been superseded by alternative one that has wider range of
applicability, yet preserving the variable-separability beyond linear 
regime \cite{b12}:  
Namely, put  
\begin{eqnarray}\label{eq18}
  \psi(z,\widehat{z}) &=& \mbox{R.P.} \;\tau \hspace{-1mm}\int^{\infty}_{-\infty}  \phi(z,\omega) \phi^\ast
  (\widehat{z},\omega) \:d\omega \nonumber \\ 
  &=& \mbox{R.P.} \;\tau \hspace{-1mm}\int^{\infty}_{-\infty}  \phi(z,\omega) \phi(\widehat{z},\widehat{\omega})
  \delta(\omega+ \widehat{\omega})  
  \:d\omega d\widehat{\omega}
\end{eqnarray}
\begin{equation}\label{eq19}
  \psi (z,\widehat{z},\widetilde{z}) = \mbox{R.P.} \;\tau^2 
 \hspace{-1mm}\int^{\infty}_{-\infty}  \phi_3 (z,\omega) \phi_3 
  (\widehat{z}, \widehat{\omega}) \phi_3 (\widetilde{z},\widetilde{\omega})
  \delta (\omega+ \widehat{\omega}+ \widetilde{\omega}) \:d\omega
  d\widehat{\omega} d \widetilde{\omega} 
\end{equation}
where $\omega$ is the variable-separation parameter having the dimension of 
the frequency,  $\tau$ is a characteristic time,
symbols($\ast$) and R.P. denote the complex conjugate and the real
part, respectively. Separated variable $\phi$ is complex, subject to 
\begin{equation}\label{eq20}
  \phi^\ast (z,\omega) = \phi(z, -\omega)
\end{equation}
as will be justified a posteriori.(See Eq.(\ref{eq22}) below.) The
closure condition is introduced in the following form 
\begin{equation}\label{eq21}
  \phi_3 = \phi
\end{equation}
It makes Eq.(\ref{eq12}) separated into two equations each for
respective points $z$ and $\widehat{z}$, in the form of complex conjugate
to each other provided that condition (\ref{eq20}) is met :
%
%
\begin{equation}\label{eq22}
  (-i\omega + \frac{\partial}{\partial t} + \myvec{v} \cdot 
\frac{\partial}{\partial  \myvec{x}} )\: \phi 
- J (z|\widehat{z})[ \phi \hat{f} + f \widehat{\phi} +
\tau \hspace{-1mm}\int^{\infty}_{-\infty}  \phi(\omega-\widetilde{\omega}) \widehat{\phi}(\widetilde{\omega}
)d  \widetilde{\omega}] = 0 
\end{equation}
If the nonlinear term of convolutional integral is deleted, the
equation degenerates to the previous case with tertiary chaos closure.

Fluid moments of (\ref{eq22}) together with (\ref{eq18}) provide equations for
turbulent transports such as Reynolds' stress and turbulent heat flux
density to obey. For actual derivation of these
equations see \cite{b13}. 
\section{Fluctuation equations in physical-plus-eddy space}
The approach described in the previous section has shed some lights in 
turbulence research. In fact, for the cases tested agreement
with experiments is satisfactory although the theory is free from any
adjustable parameters\cite{b14,b15} as contrast with existing models 
such as eddy-viscosity model. The success, however, has been
limited to cases 
where the flow geometry is governed by single variable.(Note that the
velocity fluctuations are multi-dimensional.)

We will show that a small renovation of the theory sketched in the
preceeding 
section can make turbulence with general three-dimensional geometry
tractable. It is to replace the frequency $\omega$, a scaler quantity
having appeared in Eqs.(\ref{eq22}) and (\ref{eq20}), by the wave
number \myvec{k} that 
is a vector connected to the frequency through phase velocity \myvec{c} 
by the dispersion formula 

\begin{equation}\label{eq23}
  \omega = \myvec{c} \cdot \myvec{k}  
\end{equation}
Then, new separation rule to replace (\ref{eq18}) is 

\begin{equation}\label{eq24}
  \psi(z,\widehat{z}) =
  \mbox{R.P.}\;l^3 \hspace{-1mm}\int^{\infty}_{-\infty} \phi(z,\myvec{k})\:\phi^*(\widehat{z},\myvec{k})
  \:d\myvec{k}
\end{equation}
where $l$ is the characteristic length of the macroscopic phenomenon
under consideration. Accordingly the governing equation for $\phi$ to
replace (\ref{eq22}) is written as 

\begin{equation}\label{eq25}
  \left.
    \begin{array}{l}
      \displaystyle
      i\omega\phi = \Omega(\phi)\\
      \mbox{with}\\
      \displaystyle
      \Omega(\phi) \equiv \left(\frac{\partial}{\partial t} +
        \myvec{v} \cdot \frac{\partial}{\partial \myvec{x}} \right)\phi -
      J(z|\widehat{z})\left[f\widehat{\phi} + \phi\hat{f} +
        \mbox{R.P.}\;l^3 \hspace{-1mm}\int^{\infty}_{-\infty} \phi(\myvec{k} -
        \widetilde{\myvec{k}})\widehat{\phi}(\widetilde{\myvec{k}})
        d \widetilde{\myvec{k}}\right]\\  
    \end{array}
  \right\}
\end{equation}

For practical purposes it is more convenient to separate out the
spacially periodic factor from $\phi$ by putting

\begin{equation}\label{eq26}
  \phi(z,\myvec{k}) = e^{i \myvec{k}\cdot\myvec{x}}\Phi(z,\myvec{k})
\end{equation}
and work with its amplitude $\Phi$. Owing to the fortuitous situation
that the only nonlinear term in Eq.(\ref{eq25}) has the form of
convolutional integral, the factor $e^{i\myvec{k}\cdot\myvec{x}}$
is seen 
to drop off upon substituting expression (\ref{eq26}) into 
Eq.(\ref{eq25}). Since $(\partial/\partial x_j)\phi =
e^{i\myvec{k}\cdot\myvec{x}}(\partial/\partial x_j + ik_j)\Phi$
the equation 
for $\Phi$ should preserve the form of Eq.(\ref{eq25}) with the only
substitution 

\begin{equation}\label{eq27}
  \frac{\partial}{\partial x_j} \longrightarrow \frac{\partial}{\partial x_j} +
  ik_j
\end{equation}

Another favorable property of the nonlinear integral is that it reduces 
to a simple product through Fourier transform 

\begin{equation}\label{eq28}
  \Phi(z,\myvec{k}) = \frac{1}{(2\pi l)^3}\int^{\infty}_{-\infty} 
  e^{-i \myvec{k}\cdot\myvec{s}} F(z,\myvec{s})\:d\myvec{s}
\end{equation}
such that

\begin{equation}\label{eq29}
  \int^{\infty}_{-\infty} \Phi(\myvec{k} - \widetilde{\myvec{k}})\:\widehat{\Phi}(\widetilde{
\myvec{k}}) \:d\widetilde{\myvec{k}}
  \longrightarrow F\widehat{F}
\end{equation}
Then Eq. (\ref{eq25}), with (\ref{eq26}) through 
(\ref{eq29}) taken into account, reads

\begin{equation}\label{eq30}
  \left(\frac{\partial}{\partial t} -
    c_j\frac{\partial}{\partial s_j} + v_j\partial_j\right)
  F - J(z|\widehat{z})\left[f\widehat{F} + \hat{f} F + F\widehat{F}\right] = 0
\end{equation}
where $\partial_j$ is six-dimensional operator defined by 

\begin{equation}\label{eq31}
\partial_j \equiv \frac{\partial}{\partial x_j} +
\frac{\partial}{\partial s_j}
\end{equation}

The Fourier variable $\myvec{s}$ introduced through the transform
(\ref{eq28}) has the dimension of length characterizing the size of
eddies. Therefore $\myvec{s}$ might well be called the eddy variable.

Substituting (\ref{eq26}) and (\ref{eq28}) into (\ref{eq24}) we have a 
remarkably simple formula for $\psi$ when written in terms of $F$,

\begin{equation}\label{eq32}
  \psi(z,\widehat{z}) = \frac{1}{(2\pi l)^3}\int^{\infty}_{-\infty} 
  F(\myvec{x},\myvec{s},\myvec{v})\:F(\widehat{\myvec{x}},\myvec{s} +
  \widehat{\myvec{x}} - \myvec{x},\widehat{\myvec{v}})\:d\myvec{s}
\end{equation}
Note that here all the quantities of the integrand are real in
contrast with those of (\ref{eq24}).

Derivation of fluid equations from Eq.(\ref{eq30}) is almost parallel
to its low dimensional predecessor\cite{b13}, so will not be repeated
here. Minimum necessary steps to reach the final form of fluid
equations will be given in order : First we employ the same
fluid-moment expansion for F($\myvec{x},\myvec{s},\myvec{v},t$) in terms of 
Hermite's polynomial $\cal H$ in the $\myvec{v}$-space after Grad \cite{b16};

\begin{equation}\label{eq33}
  \left.
    \begin{array}{rl}
\displaystyle
      F & = \displaystyle\frac{e^{-\xi^2/2}}{(2\pi a^2)^{3/2}m} 
        \left[q_0 + \frac{q_j}{a} \:{\cal H}_j
        + \frac{q_{jk}}{2a^2} \:{\cal H}_{jk}
        + \frac{q_{jkk}}{10a^3} \:{\cal H}_{jrr}\right]\\  
      \mbox{with}\\ 
	{\cal H}& = {\cal H }(\myvec{\xi}) \\
      \myvec{\xi}& = \myvec{w}/a = (\myvec{v} - \myvec{u}^{\dagger})/a\\
      u\sp{\dagger}\sb{j}& \equiv m_j / \rho\\
      m_j &\equiv \rho u_j + \overline{\rho' u_j'}\\
      a^2 &\equiv R_M T \quad (R_M \equiv \mbox{Avogadro no./molecular
        no.})\\ 
    \end{array}
  \right\}
\end{equation}
Note that this expansion differs from the classical 13-moment expansion in 
that $q_j \ne 0, q_{jj} \ne 0$.  
Second, we replace Eq.(\ref{eq30}) by its moment equivalents 
\begin{equation}\label{eq34}
  m\hspace{-1mm}\int^{\infty}_{-\infty}  M_\alpha[\mbox{Eq.}(\ref{eq30})]d\myvec{v} = 0
\end{equation}
where $F$ is substituted by (\ref{eq33}) and $M_\alpha$ stands for one 
of the 13 moment functions; 1,${\cal H}_j$,${\cal H}_{jj}$,${\cal
  H}_{jk}$,${\cal H}_{jkk}$. These choices of moments correspond to
equations of fluctuations of continuity, momentum, energy, stress
tensor $q_{jk}$ and heat flux density vector $q_{jkk}$, respectively. The
actual form of the equations are :
%
%
%
%
%
%
\begin{equation}\label{eq35}
  \left.
    \begin{array}{l}
      Dq_{0} + \partial_rq_r = 0\\
      \displaystyle
      Dq_{j} + \partial_rq_{jr} + \partial_jq_{40} + 
      \frac{\partial u\sp{\dagger}\sb{j}}{\partial x_r}q_r
      - \frac{1}{\rho}\frac{\partial p}{\partial x_j} q_0 = 0\\  
\vspace{-4mm}\\
      \displaystyle
      \frac{3}{2} Dq_{40} + \partial_rq_{rjj} +
      \frac{5}{2}\partial_r(a^2q_r) + \frac{\partial u\sp{\dagger}\sb{r}}
      {\partial x_r} q_{40} + 
      \frac{\partial u\sp{\dagger}\sb{j}}{\partial x_r} q_{jr} -
      \frac{1}{\rho}\frac{\partial p}{\partial x_r} q_r = 0\\ 
\vspace{-4mm}\\
      q_{40} = q_4 + a^2q_0\\
\vspace{-4mm}\\
      \displaystyle
      q_{jr} = -\mu\left[\partial_r\left(\frac{q_j}{\rho}\right) +
        \partial_j\left(\frac{q_r}{\rho}\right) -
        \frac{2}{3}\delta_{jr}\partial_k\left(\frac{q_k}{\rho}\right)\right]
      + \frac{1}{\rho}\left[q_jq_r - \frac{1}{3}\delta_{jr}{q_k}^2\right]\\
\vspace{-4mm}\\
      \displaystyle
  \hspace{10mm} -\frac{1}{\rho R_M}\frac{d\mu}{dT} 
    \left[\frac{\partial u\sp{\dagger}\sb{j}}{\partial x_r} + 
          \frac{\partial u\sp{\dagger}\sb{r}}{\partial x_j} - 
          \frac{2}{3}\delta_{jr}
          \frac{\partial u\sp{\dagger}\sb{k}}{\partial x_k}\right]q_4\\
\vspace{-3mm}\\
      \displaystyle
      q_{rjj} = -\frac{\lambda}{R_M} \partial_r\left(\frac{q_4}{\rho}\right) -
       \frac{1}{\rho R_M}\frac{d \lambda}{dT} 
       \frac{\partial T}{\partial x_r} q_4 + 
      \frac{5}{2}\frac{q_rq_4}{\rho}\\
\vspace{-4mm}\\
      \mbox{where we have defined the following symbols :}\\
\vspace{-4mm}\\
      \displaystyle
      Dq \equiv \frac{\partial q}{\partial t} 
          - c_r \frac{\partial q}{\partial s_r}
          + \partial_r (u\sp{\dagger}\sb{r} q)
    \end{array}
  \right\}
\end{equation}
In the above, $\rho$, $p$, $T$, $\mu$ and $\lambda$ denote density,
pressure, temperature,viscosity and thermal conductivity coefficients, 
respectively.

It is readily seen that Eqs.(\ref{eq35}) are exactly the same as those 
derived in ref.\cite{b13}[Eqs.(\ref{eq25}') through (\ref{eq28}')] if
the following replacements are effected,

\begin{equation}\label{eq36}
  i\omega \longrightarrow c_j\frac{\partial}{\partial s_j}
\end{equation}

\begin{equation}\label{eq37}
  \frac{\partial}{\partial x_j} \longrightarrow \frac{\partial}{\partial x_j} +
  \frac{\partial}{\partial s_j}
\end{equation}


The set of Eqs.(\ref{eq35}) describes evolution of five quantities $(q_0, 
q_j, q_4)$ subject to homogeneous boundary conditions that all $q$'s vanish 
with $|\myvec{s}| \to \infty$, on the solid surface, and wherever turbulent 
intensity is zero in the physical space. Therefore, the expected solution 
for those quantities must have the form of a solitary wave (not to be confused 
with a soliton).

Eqs.(\ref{eq35}) are equations governing
compressible turbulence. For incompressible flows $(q_0 =0)$ 
the energy fluctuation equation is decoupled, and a closed set of equations that results is

\begin{equation}\label{eq39}
  \partial_jq_j = 0
\end{equation}

\begin{equation}\label{eq40}
  (D - \nu\partial\sp{2}\sb{r})q_j + \partial_jq_{40} + \frac{\partial 
    u_j}{\partial x_r}q_r + \frac{q_r}{\rho}\partial_rq_j = 0
\end{equation}
where $\nu$ is the kinematic viscosity, $\partial_j$ and $D$ have been 
defined by (\ref{eq31}) and (\ref{eq35}), respectively. 
It is readily seen that these
equations represent the equation of continuity and the Navier-Stokes
equation generalized to 6D space $(\myvec{x},\myvec{s})$. In fact, if
we suppress the eddy variables$(\partial /\partial s_j = 0)$ this set
degenerates to the usual Navier-Stokes equation for velocity $u_j +
\rho^{-1}q_j$ and pressure $p + q_{40}$. If further, nonlinear terms
are neglected in Eq.(\ref{eq40}) and parallel flow$(u_j =
\delta_{j1}u(x_2))$ is assumed, they reduce to the Orr-Sommerfeld
equation to govern $\rho^{-1}q_2$ as it should.
\section{Correspondence rule; relationships between solitary-wave functions 
and observables}
It should be remarked that variables $q$'s do not correspond to any
turbulent fluctuations that are tangible to macroscopic sensors.They
are shown to be related to fluctuation correlations of turbulent
quantities through 
the following deduction: From (\ref{eq7}), (\ref{eq9})
and (\ref{eq1}) we
have expression for instantaneous density fluctuation that is an observable; 

\begin{equation}
  \label{eq41}
\rho'=m\hspace{-1mm}\int^{\infty}_{-\infty}  f^{'} d \myvec{v}  
\end{equation}
Similarly, by subtracting (\ref{eq8}) from (\ref{eq10}), we have
\[
\rho u'_j + \rho' u_j + (\rho'u'_j-\overline{\rho' u'_j})=m\int^{\infty}_{-\infty} v_j f'd\myvec{v}
\]
This expression, upon substituting (\ref{eq41}) for $\rho'$, reduces to
\begin{equation}
\rho u'_j= m\int^{\infty}_{-\infty} w_j f' d\myvec{v} + O(f'^3)
\end{equation}
with
\[
O(f'^3)\equiv \rho^{-1}\rho'\overline{\rho'u'_j} - (\rho'u'_j-\overline{\rho'u'_j})
\]
%
%
from which we have, utilizing definition (\ref{eq6}) for \(\psi\) and neglecting terms of $O(f'^4)$
 
\begin{equation}
  \label{eq43}
 \rho\widehat\rho\: \overline{u_j'\:\widehat u_l'} = m^2\hspace{-1mm}\int^{\infty}_{-\infty}  w_j 
 \:\widehat w_l \:\psi(z,\widehat z)
 \:d \myvec{v}d\widehat{\myvec{v}} 
\end{equation}
Furthermore, by substituting (\ref{eq32}) for \(\psi\) and then
(\ref{eq33}) for $F$, and carrying out integration
over(\(\myvec{v},\widehat{\myvec{v}}\)) with orthonormal property of the
Hermite polynomials incorporated, the following relationship results;

\begin{equation}
  \label{eq44}
  \rho\widehat\rho\: \overline{u_j' \:\widehat u_l'} = \frac{1}{(2\pi
    l)^3}\int^{\infty}_{-\infty}  q_j(\myvec{x},\myvec{s})\:q_l(\widehat{\myvec{x}},\myvec{s}+
  \widehat{\myvec{x}}- \myvec{x})\:d\myvec{s}
\end{equation}
Turbulent intensity or Reynolds' stress is then given by putting
\(\widehat{\myvec{x}}=\myvec{x}\);

\begin{equation}
  \label{eq45}
  \overline{u_j'\:u_l'}=\frac{1}{\rho^2(2\pi l)^3}
  \int^{\infty}_{-\infty}  q_j(\myvec{x},\myvec{s})\:q_l(\myvec{x},\myvec{s})\:d\myvec{s}
\end{equation}
a relationship expressing the observable turbulence intensities by an 
integral operation of the wave function. 
It is through this relationship that the Reynolds averaged
Navier-Stokes equation is coupled with Eqs.(\ref{eq35}) that govern
$q$'s standing on the r.h.s. of (\ref{eq45}).

In a similar fashion the wave function representing temperature fluctuation 
$T^{'}$ can be derived as follows: Since the ideal gas law $\underline{p} 
= R_M \underline{\rho} \underline{T}$ holds for instantaneous variables 
\cite{b5}, we have 
\begin{eqnarray}
R_M \rho T^{'} &=& p' - R_M T \rho' \nonumber \\
 &=& \frac{m}{3} \int^{\infty}_{-\infty}  w_j^2 f^{'} d \myvec{v} - a^2 m \int^{\infty}_{-\infty}  
 f^{'} d \myvec{v} \nonumber \\
 &=& \frac{m a^2}{3} \int^{\infty}_{-\infty}  {\cal H}_{jj} f^{'} d \myvec{v}, \nonumber
\end{eqnarray}
Accordingly, from (\ref{eq32})
\begin{eqnarray}
R_M^2 \rho \widehat{\rho}\; \overline{T^{'} \:\widehat{T^{'}}} &=& 
 (\frac{m}{3})^2 a^2 \:\widehat{a}^2 \int^{\infty}_{-\infty}  {\cal H}_{jj} \widehat{{\cal H}}_{ll}
 \:\psi \:d \myvec{v} d \widehat{\myvec{v}} \nonumber \\
  &=& \frac{1}{(2 \pi l)^3 3^2} \int^{\infty}_{-\infty}  q_{jj} (\myvec{x}, \myvec{s}) 
  \:q_{ll}(\widehat{\myvec{x}}, \myvec{s} + \widehat{\myvec{x}} - \myvec{x})
  \:d \myvec{s} \nonumber 
\end{eqnarray} 
where we have employed Hermite expansion (\ref{eq33}). Thus we see that 
$q_4 = (1/3) q_{jj}$ is the wave function to be responsible for the temperature
 fluctuation $R_M \rho T^{'}$.

Summarizing, the  following {\it correspondence rule} holds between {\em 
untangible} wave function \(q_\alpha\) and
corresponding {\em observable} fluctuation \(A^{'}_\alpha\);

\begin{equation}
  \label{eq46}
  q_\alpha=
  \left(
    \begin{array}{c}
      q_0\\
      q_j\\
      q_4 = \frac13 q_{jj}\\
      q_{40}
    \end{array}
  \right),\ 
  A^{'}_\alpha=
  \left(
    \begin{array}{c}
      \rho'\\
      \rho u_j'\\
      \rho R_M T^{'}\\
      p'
    \end{array}
  \right)
\end{equation}
where the fourth quantity \(q_{40}\) in the column of \(q_\alpha\) is
linearly dependent on \(q_0\) and \(q_4\).(See Eq.(\ref{eq35}).) 
They are related 
to each other through the following fluctuation correlation formula ;
\begin{equation}
  \label{eq47}
  \overline{A^{'}_\alpha \widehat{A}^{'}_\beta}=
  \frac{1}{(2\pi l)^3 }\int^{\infty}_{-\infty}  q_\alpha(\myvec{x},\myvec{s})
  \:q_\beta(\widehat{\myvec{x}},\myvec{s}
  +\widehat{\myvec{x}}- \myvec{x})d\myvec{s}
\end{equation}
In particular, for example, turbulent heat flux density is given as
follows, using (\ref{eq46}) and (\ref{eq47}) with
\(\widehat{\myvec{x}}=\myvec{x}\);

\begin{equation}
  \label{eq48}
  c_p\rho^2\:\overline{T^{'} u_j'}=\frac{5}{2}\int^{\infty}_{-\infty}  q_4 \:q_j d\myvec{s}
\end{equation}
where \(c_p\) is the specific heat under constant pressure.
\section{Reformulation within phenomenologies}
Once the correspondence rule  (\ref{eq46}) has been established we are
able to reconstruct Eq.(\ref{eq35}) using phenomenologies alone.This
is what is anticipated because the present (Boltzmann) formalism is
the averaged version of the Klimontovich formalism describing 
\(A^{'}_\alpha\) directly, where the identity of phenomenologies 
and first-principle approach is warranted \cite{b5}.

To effect this we shall base on the assertion that turbulent
compressible flow of inert gas is governed by the following set of equations:

\begin{equation}
  \label{eq49}
  \Lambda_0\equiv\frac{\partial\underline{\rho}_j}{\partial t}
  +\frac{\partial\underline{m}_r}{\partial x_r}=0
\end{equation}

\begin{equation}
  \label{eq50}
  \Lambda_j\equiv\frac{\partial\underline{m}_j}{\partial t}+ 
  \frac{\partial}{\partial x_r}
  \left[\frac{\underline{m}_j\underline{m}_r}{\underline{\rho}}
  +\underline{p}\delta_{jr}
  +\left(\underline{p}_{jr}\right)_{NS}\right]=0
\end{equation}

\begin{equation}
  \label{eq51}
  \Lambda_4\equiv\frac{\partial}{\partial t}
  \left(\underline{E}+\frac{\underline{m}^2_j}{2\underline{\rho}}\right)
 +\frac{\partial}{\partial x_r}
  \left[\frac{\underline{m}_r}{\underline{\rho}}\left(\underline{H}+
  \frac{\underline{m}^2_j}{2\underline{\rho}}\right)
 +\frac{\underline{m}_j}{\underline{\rho}}
  \left(\underline{p}_{jr}\right)_{NS}+\left(\underline{Q}_r\right)_{F} 
  \right]=0
\end{equation}
%
where
\begin{equation}
  \label{eq52}
  \underline{m}_j\equiv\underline{\rho}\underline{u}_j
\end{equation}

\begin{equation}
  \label{eq53}
  \left.
    \begin{array}{ll}
  \underline{E} \equiv \underline{\rho}\underline{e}=\underline{p}/\left(\gamma-1\right) & (\gamma; \mbox{specific heats ratio})\\
\vspace{-3mm}\\
  \underline{H} \equiv \underline{\rho}\underline{h}=\underline{p}\left[\gamma/\left(\gamma-1\right)
  \right]
  \end{array}
\right\}
\end{equation}
\begin{equation}\label{eq54}
\left.
	\begin{array}{lcl}
  \left(\underline{p}_{jr}\right)_{NS}&\equiv&
\displaystyle
  - \underline{\mu}\left[\frac{\partial}{\partial x_r}
   \left(\frac{\underline{m}_j}{\underline{\rho}}\right) +
   \frac{\partial}{\partial x_j}
  \left(\frac{\underline{m}_r}{\underline{\rho}}\right)\right]-
   \frac{2}{3} \delta_{jr}
   \frac{\partial}{\partial x_k}
   \left(\frac{\underline{m}_k}{\underline{\rho}}\right)\\
\vspace{-3mm}\\
  \left(\underline{Q}_r\right)_{F}&=&
\displaystyle
  -\frac{\underline{\lambda}}{R_M}
  \frac{\partial}{\partial x_r}
  \left(\frac{\underline{p}}{\underline{\rho}}\right)
\end{array}\right\}
\end{equation}
denote, respectively, the mass flux density, the internal energy and
the enthalpy per unit of volume (the lower case letter refers to 
{\em specific} quantities), the Navier-Stokes law for stress
tensor and the Fourier's law for the heat flux density.

It has been shown that plain (Reynolds) average of this set of equations

\begin{equation}
  \label{eq54a}
  \overline{\Lambda}_\alpha=0\ \ \ (\alpha=0,j,4)
\end{equation}
is in exact coincidence with the first principle deduction for
monatomic gases\(\left(\gamma=\frac{5}{3}\right)\)\cite{b17}.
Written explicitly, they are 
\begin{equation}
  \label{eq54b}  
  \overline{\Lambda}_0\equiv\frac{\partial\rho}{\partial t}+
  \frac{\partial m_r}{\partial x_r}=0
\end{equation}

\begin{equation}
  \label{eq55}
  \left.
    \begin{array}{rcl}
      \overline{\Lambda}_j&\equiv&
      \displaystyle\frac{\partial m_j}{\partial t} +
      \frac{\partial}{\partial x_r}
      \left(\frac{m_jm_r}{\rho}+p\delta_{jr}+p_{jr}\right)=0\\
\vspace{-3mm}\\
      p_{jr}&=&\left(p_{jr}\right)_{NS}+\rho\:\overline{u_j' u_r'}\\
\vspace{-3mm}\\
      \left(p_{jr}\right)_{NS}&=&\left(\overline{\underline{p}}_{jr}
      \right)_{NS} 
    \end{array}
 \right\}
\end{equation}

\begin{equation}
  \label{eq56}
  \left. 
    \begin{array}{rcl}
      \overline{\Lambda}_4 &\equiv& \displaystyle\frac{\partial}{\partial t}
      \left(E+\frac{m^2_j}{2\rho}\right) + \frac{\partial}{\partial x_r}
      \left( \frac{m_r}{\rho} H+\frac1\rho m_r m^2_j +
        \frac{m_j}{\rho} p_{jr}+Q_r\right)=0\\
\vspace{-3mm}\\
      Q_r &=& \left(Q_r\right)_F+\rho c_p\: \overline{T^{'} u_r'}\\
\vspace{-3mm}\\
      \left(Q_r\right)_{F} &=& \left( \overline{\underline{Q}}_r \right)_{F}
    \end{array}
  \right\}
\end{equation}
A few remarks are in order :
\begin{itemize}
\item[i)] These equations are exact to \(O(A^{'2})\).
\item[ii)] They are 
written in terms of quantities that are {\em proportional to
the density}, for example, mean mass-flux density
$$
m_j\equiv\overline{\underline{m}}_j=\rho u_j+\overline{\rho' u_j'}
$$
to replace mean fluid velocity \(u_j\), also mean internal energy $E$ per
unit of volume
$$
E\equiv\overline{\underline{\rho}\:\underline{e}}=\frac{1}{\gamma-1}
\left(p+R_M\overline{\rho'\:T^{'}}\right)
$$
to replace the specific internal energy $e$.
\item[iii)]Item ii) is the key that enables to express
  Reynolds-averaged equations (\ref{eq55})and (\ref{eq56}) in
  {\it compressibility invariant} forms, in other words, {\em single} term
  turbulence correction to  each of stress \(p_{jr}\) and heat flux
  density \(Q_r\) suffices even under presence of 
  density fluctuation such as turbulent combustion .
\item[iv)]Otherwise, lengthy additional terms for turbulence correction
  would appear, or we would need the so-called mass averaging
  (say $\widetilde{\underline{u}}_j,$ for instance) that suffers from a
  conceptual difficulty \(\left(\widetilde{u}'_{j}\neq0 \right)\) in
  processing experimental data.
\end{itemize}

Next step, the main issue of this section, is to show the identity of Eqs. 
(\ref{eq35}) with the
following phenomenological equations 

\begin{equation}
  \label{eq55b}
  \overline{A^{'}_\alpha\widehat{\Lambda}_\beta
    +\widehat{A}^{'}_\beta\Lambda_\alpha}=0, \ 
  \left(\alpha,\beta ; 1,j,4 \right)
\end{equation}
where \(A^{'}_\alpha\)is defined in (\ref{eq46}). These equations
consist of terms of double $(O(A^{'2}))$ and triple 
\((O(A^{'3}))\) correlations for which we employ the
separation rule exactly parallel to those of the previous section 
[Eq.(\ref{eq24}) through Eq.(\ref{eq32})]:\\
Put
\begin{equation}
  \label{eq56a}
  \left.
    \begin{array}{rl}
      \overline{A^{'}_\alpha \widehat{A}^{'}_\beta}=&
\displaystyle
      \mbox{R.P.}\:l^3 \int^{\infty}_{-\infty}  g_\alpha\left(\myvec{x},\myvec{k}\right)
      g_\beta\left(\widehat{\myvec{x}},\myvec{k}\right)^\ast d\myvec{k}\\
\vspace{-3mm}\\
      \overline{A^{'}_\alpha \widehat{A}^{'}_\beta\widetilde{A}^{'}_\gamma}=&
\displaystyle
      \mbox{R.P.}\:l^6\int^{\infty}_{-\infty}  g_\alpha\left(\myvec{x},\myvec{k}\right)
      g_\beta\left(\widehat{\myvec{x}},\myvec{k}'\right)^{\ast}
      g_\gamma (\widetilde{\myvec{x}},\myvec{k}-\myvec{k}')^{\ast}
      d \myvec{k} d\myvec{k}' \\
\vspace{-2mm}\\
      =&\displaystyle
      \mbox{R.P.} \:l^6\int^{\infty}_{-\infty}  g_\alpha (\myvec{x}, \myvec{k}')\:
      g_\beta (\widehat{\myvec{x}}, \myvec{k})^{\ast} \:
      g_\gamma (
      \widetilde{\myvec{x}}, \myvec{k}-\myvec{k}')\:d\myvec{k}d \myvec{k}' \\
\vspace{-3mm}\\
      \mbox{with} \ & g_{\alpha} (\myvec{x},-\myvec{k})
      = g_{\alpha}(\myvec{x},
      \myvec{k})^{\ast}
    \end{array}
  \right\}
\end{equation}
and substitute into Eq.(\ref{eq55b}), then we are led to the equation
that allows for the separation of variables :

\begin{equation}
  \label{eq57}
  \begin{array}{c}
    \displaystyle\int^{\infty}_{-\infty}  d\myvec{k}\: g_\alpha
    \left(\widehat{g}_\beta\right)^\ast
    \left[\frac{\Lambda^\dagger_\alpha}{g_\alpha}
      +\left(\frac{\widehat{\Lambda}^\dagger_\beta}{\widehat{g}_\beta}
    \right)^{\hspace{-1mm}\ast} \;\right]=0\\
    \hspace{18mm} \Vert \hspace{11mm}  \Vert\\
    \hspace{19mm} i\omega \hspace{5mm} -i\omega
  \end{array}
\end{equation}
In the above \(\Lambda^\dagger_\alpha\) is the
fluctuating part of \(\Lambda_\alpha\) in which \(A^{'}_\alpha\) is
replaced with \(g_\alpha\), also \(A^{'}_\alpha A^{'}_\gamma\) 
with convolution integral \(\displaystyle\int^{\infty}_{-\infty} 
g_\alpha\left(\myvec{k'}
\right)g_\gamma(\myvec{k}-\myvec{k'})d\myvec{k'}\), and $\partial /
\partial x_{r}$ with $\partial / \partial x_r + ik_r$. The separated equation 
thus obtained, namely,

\begin{equation}
  \label{eq58}
  -i\omega g_\alpha+\Lambda^\dagger_\alpha=0
\end{equation}
is then rewritten in Fourier-analyzed form after
\begin{equation}\label{eq59}
  g (\myvec{x,k})= \frac{1}{(2 \pi l)^3} \int^{\infty}_{-\infty}  e^{-i
    \myvec{k} \cdot \myvec{s}}q(\myvec{x,s})\:d\myvec{s}
\end{equation}
to lead to the same equations as Eqs.(\ref{eq35}) except for the
energy fluctuation equation which takes the form,
\begin{equation}\label{eq60}
  \frac{1}{\gamma -1} Dq_{40} + \partial_r q_{rjj} +
  \frac{\gamma}{\gamma -1}  \partial_r (a^2 q_r) + \frac{\partial
    u_r^\dagger}{\partial x_r} q_{40} + \frac{\partial
    u_j^\dagger}{\partial x_r} q_{jr} - \frac{1}{\rho} \frac{\partial
    p}{\partial x_r} q_r = 0
\end{equation}
This equation reduces in the case of monatomic gases $(\gamma=5/3)$ to
the third of Eqs.(\ref{eq35}) as it should.
\section{A solitary-wave solution for mixing layer turbulence}
A preliminary computation checking whether the present approach is 
physically sound has been carried out for turbulent mixing shear layer 
of an incompressible flow \cite{b18}. Eqs.(\ref{eq39}) and
(\ref{eq40}) are employed assuming the average flow profile
$[u(\eta),v(\eta), 0]$ with $\eta \equiv x_2 / \alpha x_1$ as prescribed. 
The flow is self-similar in this sense as confirmed by experiment \cite{b19}, 
which is an indicative of molecular viscosity playing no roles in the 
equation. (See ref.\cite{b18}.) Then 
we have the following set of equations :
\begin{equation}\label{eq61}
  \left.
    \begin{array}{l}
      \partial_1 q_1 + \partial_2 q_2 + \partial q_3 / \partial s_3 =
      0 \\
      \underline{NL} q_1 + \partial_1 q_{40} - \alpha \eta u'q_1 +
      u'q_2=0\\ 
      \underline{NL}q_2 + \partial_2 q_{40} - \alpha \eta v'q_1 +
      v'q_2=0\\ 
      \underline{NL}q_3 + \partial q_{40} / \partial s_3 = 0
    \end{array}
  \right\}
\end{equation}
where $u' \equiv du / d \eta $, and
\begin{equation}\label{eq62}
  \left.
    \begin{array}{rcl}
      \partial_1 &\equiv& (1- \alpha s_1) \partial/\partial s_1 - 
      \alpha (\eta \partial/\partial \eta + s_2
      \partial/\partial s_2 +  s_3 \partial/\partial
        s_3) \\
      \partial_2 &\equiv& \partial/\partial \eta +
      \partial/\partial s_2  \\
      \underline{NL} &\equiv& \partial/\partial t - c\:
      \partial/\partial s_1 + (u + q_1)\partial_1 + (v + q_2)
      \partial_2 + q_3 \partial/\partial s_3 
    \end{array}
  \right\}
\end{equation}
with $\myvec{s}$ redefined using the shear mixing layer
thickness $l = \alpha 
x_1$, namely, $\myvec{s}/l \to \myvec{s}$.

The set of equations has five independent variables $(\myvec{s}, \eta, t)$, 
therefore no existing tools are immediately available. 
At this preliminary stage of checking physical soundness of the proposed 
approach it is advisable to suppress variable $s_2$ by assuming
$\partial / \partial 
\eta >> \partial / \partial s_2$. 
The set of equations is solved for arbitrary chosen initial
values for $q_\alpha$ and boundary conditions 
\begin{equation}\label{eq63}
  \left.
    \begin{array}{l}
      q_j \to 0\;,\ (j=1,2,3) \ \mbox{as} \ |\myvec{s}|, |\eta| \to
      \infty \\
\vspace{-5mm}\\
q_{40} \to 0 \ \mbox{as} \ |\myvec{s}| \to \infty\;,\; \partial
      q_{40} / \partial \eta \to 0 \ \mbox{as} \ \eta \to \infty
    \end{array}
  \right\}
\end{equation}

The form of solution to be expected from 
these homogeneous boundary conditions must be a solitary wave generated by the
shearing motion and kept sustained by nonlinearity.

In Figs.1 are shown such standing waves that build up and reach steady
state with elapse of time for different choices of the wave speeds ;
(a) experimentally observed one $c=u_0=[u(\infty)+u(- \infty)]/2$, and (b)
Taylor's hypothesis $c=u(\eta)$. Reynolds' stress
[Eq.(\ref{eq45})] is calculated using this solution and compared with 
existing experiment \cite{b19} in Fig.2. Agreement is more than
reasonable considering that the theory involves no empirical constants 
to fit with experiments.
\section{Comparison with classical statistical theory and current turbulence models}
\begin{center}
{\it Consistency with classical statistical theory\/}
\end{center}
K\'arm\'an and Howarth \cite{b20}, the founders of classical
statistical theory of turbulence, have derived Eq.(\ref{eq55b})
 for incompressible flows $(\alpha, \beta ; 0,1,2,3)$ 
correctly on intuitive basis, with no reference to first
principles. Obviously these 6-D equations are not tractable in this
form, so homogeneity/isotropy assumptions 
 have been introduced. Here
we have employed the method of separation of variables, thereby
 the classical limitation on flow geometry is eliminated. 
It is effected, however, at the expense of introducing
additional independent variable $\myvec{s}$ in the Navier-Stokes
equation as we have seen through Eqs.(\ref{eq39}) and (\ref{eq40}) . 

The classical assumptions of homogeneity and isotropy are often referred to as an oversimplification of reality. In fact, also here, this model is shown to lead to an unphysical solution : For the equation of continuity
\begin{equation}
\frac{\partial q_j}{\partial s_j} =0
\end{equation}
which is the homogeneous version ($\partial /\partial x_j =0$) of Eq.(\ref{eq39}), coupled with the isotropy requirement(Robertson's theorem \cite{b21})
\begin{equation}
q_j = s_j q(s)\;,\;\;\;(s\equiv |\myvec{s}|)
\end{equation}
gives
\begin{equation}
3\:q + s \frac{d q}{d s}=0
\end{equation}
Solution of this equation, namely, $q\sim s^{-3}$ causes the integral (\ref{eq45}) for turbulence intensity to diverge. To be noted is the fact that this is a direct consequence of the equation of continuity, a kinematical relationship universally valid, therefore independent of any closure condition employed.
\begin{center}
{\it Inconsistency with $k-\epsilon$ models\/}
\end{center}

We have seen that formulation in 6-D space
$(\myvec{x},\widehat{\myvec{x}})$ [Eq.(\ref{eq55b})] is the only one that is consistent with
Liouville's equation, namely, with first principles of nonequilibrium
statistical mechanics. Turbulent transports such as (\ref{eq45}) 
and (\ref{eq48}) that are quantities in 3-D 
space are obtained from the solution in the 6-D space by putting
$\widehat{\myvec{x}} = \myvec{x}$\ {\em after the equations have been
  solved\/}. Suppressing variables {\it in the equations} as is often
employed is a pathological process. The following simple example would 
help extract what is meant by this trivial-looking warning : Let a steady-state
temperature distribution of a three-dimensional body, say, a column with 
rectangular cross-section in the $x$-$y$ plane be asked, but the information 
only on the diagonal
plane $x = y$ is required. Needless to say that proper process is to
work with 3-D Laplace equation for the solution $T(x,y,z)$, and put
$x=y$ to have  $T(x,x,z)$. 
The improper process mentioned here corresponds to solving {\it pathological}
 equation 
$(2\partial^2/\partial x^2 + \partial^2/\partial z^2)T=0$.

Majority of turbulence models currently prevailing do {\em not} concur, in
the simplest case of the homogeneous and isotropic turbulence, with
the K\'arm\'an-Howarth theory as a consequence of the
hasty reduction in independent variables as sketched here.

Summarizing, the proposed approach shares the common basis with
$k-\epsilon$ models only at the lowest level of description, namely, the
Reynolds-averaged Navier-Stokes equation, but differs substantially at
next level of the Reynolds' stress equation.

\section{Concluding remarks}

On the basis of the Boltzmann formalism, namely, nonequilibrium
statistical mechanics designed for turbulence, two sets of equations
are derived to comprise a closed set. The one is the group of
equations governing Reynolds-averaged fluid quantities, and the other
is the variable-separated version of the fluctuation-correlation
equations that reduces to the classical K\'arm\'an-Howarth equation in
the homogeneous and isotropic limit. The two sets of equations are
coupled through turbulent transports such as Reynolds' stress and
turbulent heat-flux density.

The key role of the first principle lies in that \\
(i) \ for the Reynolds averaged equations, it gives a firmer basis than 
via phenomenologies, whereas, \\
(ii) \ for the turbulent-transport equations, it reveals a hidden 
kinematical prerequisite for any turbulence governing equations to 
fulfill, thereby enabling to single out the unique form.
It is also shown that once their unique form has been established
the whole sets of equations are able to be
reconstructed a posteriori using phenomenologies alone. 

The latter set of equations is converted into a variable-separated form, 
leading to those governing solitary-wave function through which all the 
turbulent-transport properties are calculated.

The equations are solved for a self-similar turbulent mixing layer,
leading to a solitary-wave type solution in the physical-plus-eddy
space. Reynolds stress is obtained through a simple integration of the 
solution over the eddy space in a form free from any empirical 
parameters, yet showing satisfactory agreement with existing experiment.
\clearpage
\section*{Figure captions}

Fig.1 \
Solitary wave $q_1(s_1,s_3,0)$ on the plane of $\eta=0$ for two choices of
      the phase velocity $c$ ; 
\begin{itemize}
\item[a)] $c = u_0$ , where $u_0 = [u(\infty) + u(-
        \infty)]/2$ is the propagation velocity of eddies as
        observed by flow visualization, 
\item[b)] $c=u$ \ (Taylor's hypothesis).\\
\end{itemize}
Fig.2 \ 
Reynolds' stress as calculated from Eq.(40) [18] using 
 the solitary-wave solutions (Figs.1a and 1b), and compared with
 existing experiment[19].

\section*{References}
\begin{thebibliography}{99}
\bibitem{b1} S. Chapman, {\em The Mathematical theory of
    Nonuniform Gases} (Cambridge Univ. Press, 1939)
\bibitem{b2} D. Enskog, {\em Kinetische Theorie der Vorgange in
    messig verd\"unnten Gasen}(Dissertation, Uppsala, 1917)
\bibitem{b3} I. Prigogine, {\em Physica}, {\bf 15}, 273 (1949)
\bibitem{b4} K. R. Sreenivasan and C. Meneveau, {\em J. Fluid Mech.}, {\bf 173},
  357 (1986)
\bibitem{b5} S. Tsug\'e, {\em First Principle Basis of the Direct
    Numerical Simulation for Turbulence of Inert and Reactive Gases}
  (e-print archive, chao-dyn/9712020) (1997)
\bibitem{b6} Yu. L. Klimontovich, {\em The Statistical Theory of
    Nonequilibrium Processes in a Plasma} (Cambridge Massachusetts,
  MIT Press,1967)
\bibitem{b7} N. N. Bogoliubov, {\em Problems of Dynamical Theory in
    Statistical Physics} AFCRC-TR-59-235 (1959)
\bibitem{b8} K. R. Sreenivasan, {\em Annual Review of Fluid
    Mechanics} {\bf 23} (Palo Alto, Annual Rev.Inc,1991) p.539
\bibitem{b9} S. Tsug\'e and K. Sagara, {\em J.Statist. Phys.},{\bf
12},403 (1975)
\bibitem{b10} S. Tsug\'e, {\em Phys. Fluids}, {\bf 17}, 22 (1974)
\bibitem{b11} K. Sagara and S. Tsug\'e, {\em Phys. Fluids}, {\bf 25}, 1970
(1982)
\bibitem{b12} S. Tsug\'e, {\em Phys. Lett.}, {\bf A70}, 266 (1979)
\bibitem{b13} S. Tsug\'e, K. Sagara and S. Kodowaki, {\em Inernational
    Symposium on Rarefied Gasdynamics} (Univ.Tokyo Press,1984) p.19
\bibitem{b14} K. Ishibashi,  Ph D Thesis (Inst. Eng. Mech. Univ.Tsukuba) (1991)
\bibitem{b15} S. Tsug\'e and S. Ogawa, {\em Turbulent and Molecular
    Processes in Combustion} (The Sixth Toyota Conference, Amsterdam,
    Elsevier) p.35 (1993)
\bibitem{b16} H. Grad, {\em Commun. Pure Appl. Math}, {\bf 2}, 1 (1949)
\bibitem{b17} S. Tsug\'e and K. Sagara, {\em Phys. Fluids}, {\bf 19},
1478 (1976)
\bibitem{b18} B. Bai, Ph D Thesis (Inst. Eng. Mech. Univ.Tsukuba) (1995)
\bibitem{b19} J. H. Bell and R. D. Mehta, {\em AIAAJ}, {\bf 28}, 2034 (1990)
\bibitem{b20} Th. von K\'arm\'an and L. Howarth, {\em Proc. Roy. Soc.
London}, {\bf A164}, 192 (1938)
\bibitem{b21} H. P. Robertson, {\em Proc. Camb. Phil. Soc.}, {\bf 36},
209 (1940)
\end{thebibliography}
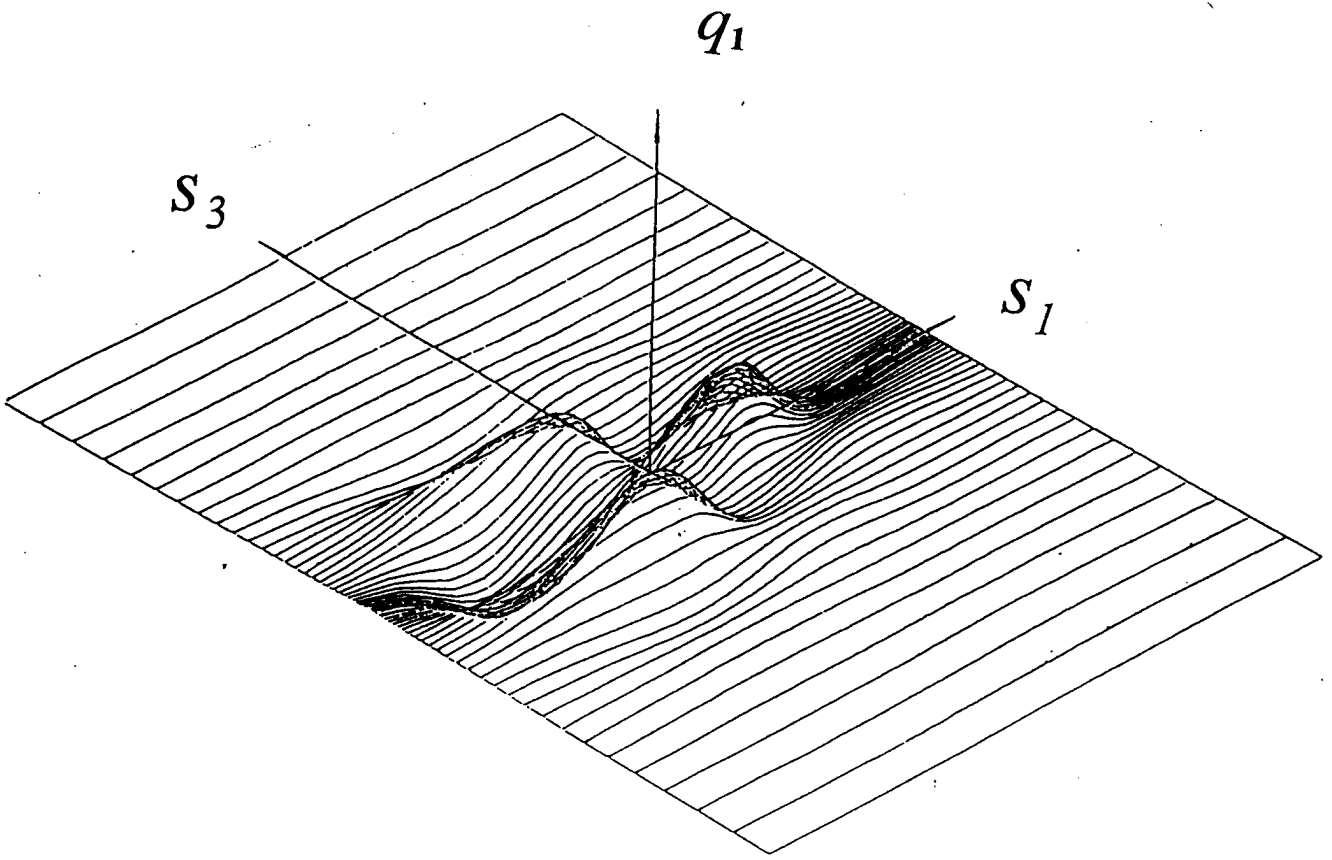
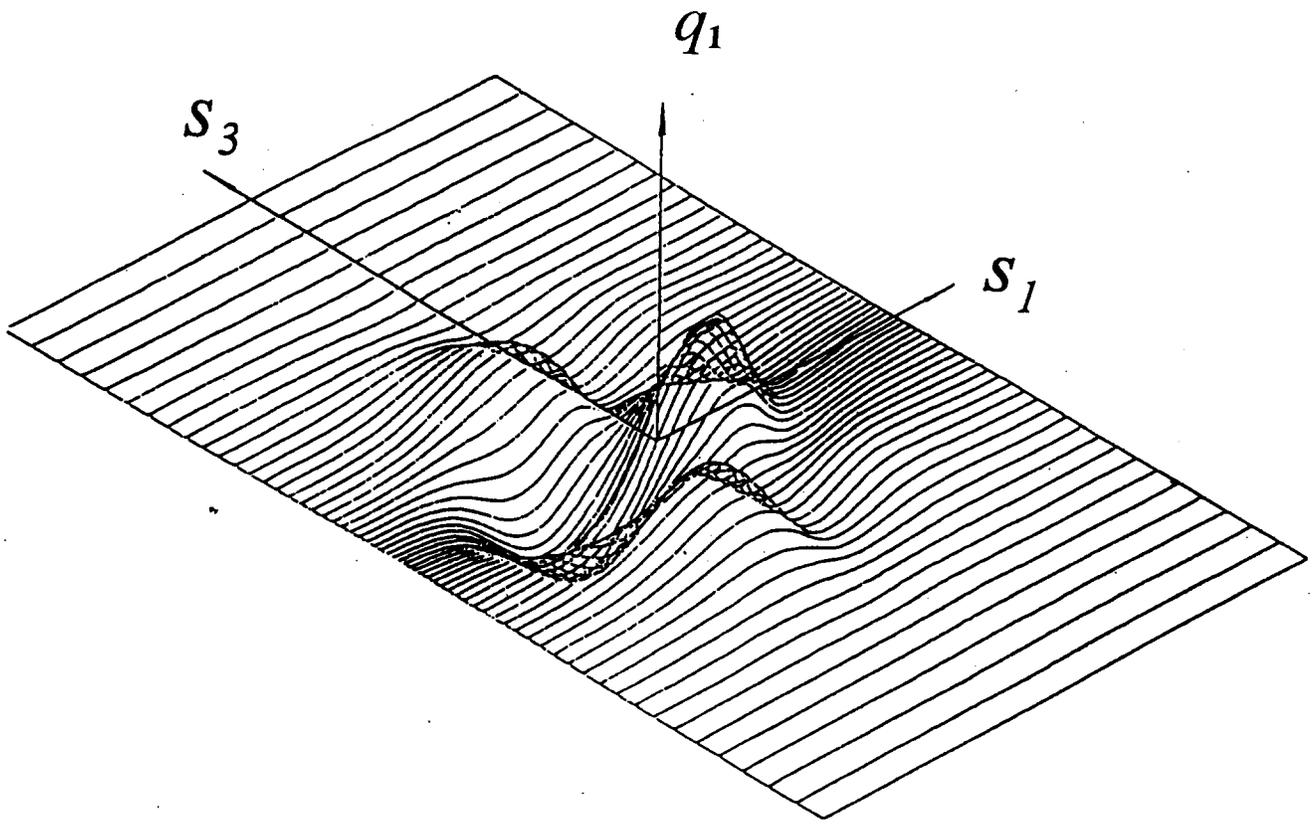
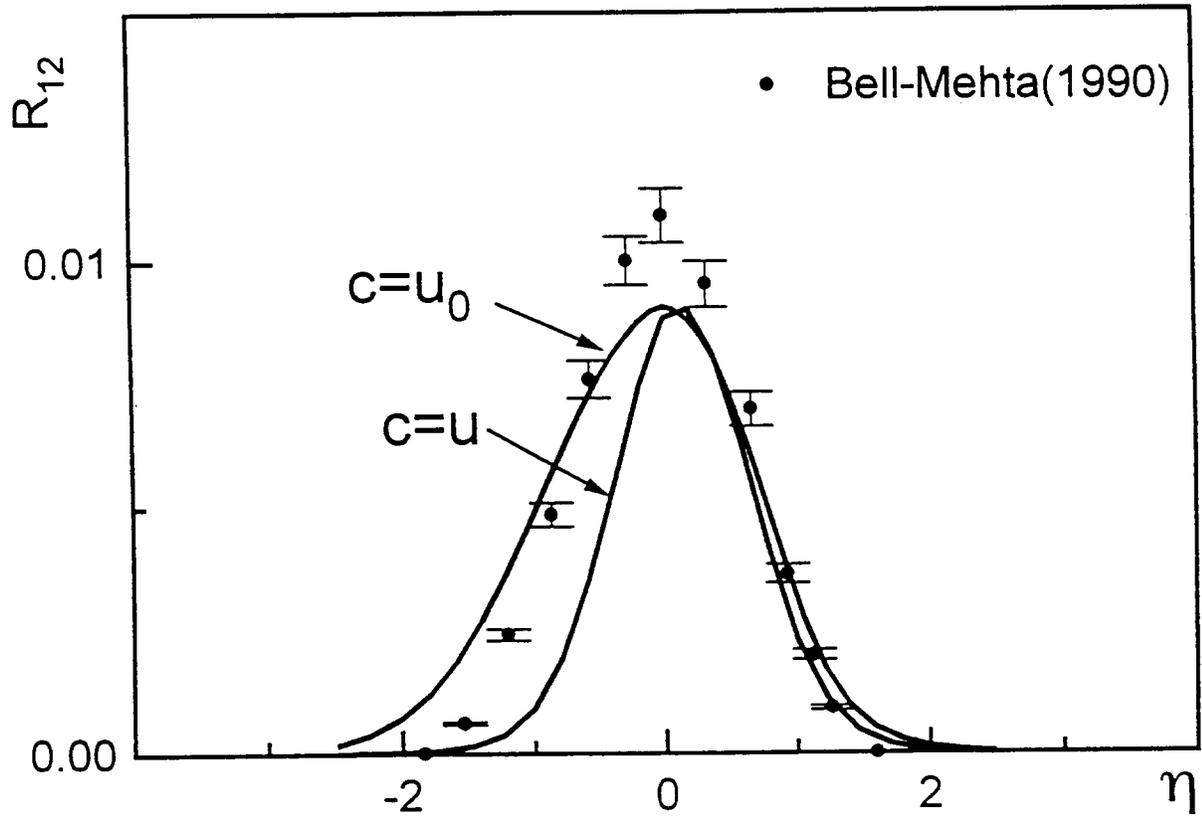
\end{document}